# BigDAWG Polystore Release and Demonstration

Kyle O'Brien[1], Vijay Gadepally[1,2], Jennie Duggan[4], Adam Dziedzic[3], Aaron Elmore[3], Jeremy Kepner[1], Samuel Madden[2], Tim Mattson[5], Zuohao She[4], Michael Stonebraker[2]

[1]MIT Lincoln Laboratory     [2]MIT CSAIL     [3]Univ. of Chicago     [4]Northwestern University     [5]Intel Corp.

**Abstract** -- The Intel Science and Technology Center for Big Data is developing a reference implementation of a Polystore database. The BigDAWG (Big Data Working Group) system supports "many sizes" of database engines, multiple programming languages and complex analytics for a variety of workloads. Our recent efforts include application of BigDAWG to an ocean metagenomics problem and containerization of BigDAWG. We intend to release an open source BigDAWG v1.0 in the Spring of 2017. In this presentation, we will demonstrate a number of polystore applications developed with oceanographic researchers at MIT and describe our forthcoming open source release of the BigDAWG system.

## I. Introduction

Polystore systems are of great interest to researchers in many diverse fields [1]. Polystore systems largely support "many sizes" of database engines [2] and a multitude of programming languages. The BigDAWG system [3] is a reference implementation of a polystore database. The BigDAWG architecture consists of four distinct layers: database and storage engines; islands; middleware and API; and applications. We have extensively described the BigDAWG architecture and specific components of the middleware in our previous publications [4,5,6,7]. In our previous work, we have described the BigDAWG system applied to the MIMIC II medical dataset [8]. Most recently, we have been working with the Chisholm Lab at MIT (https://chisholmlab.mit.edu/) to help with the analysis of ocean metagenomic data.

Our recent BigDAWG efforts have been two-fold: 1) developing a releasable version of the BigDAWG system and 2) developing applications that benefit from a polystore solution. In addition, there continue to be interesting developments in the BigDAWG middleware itself.

In this article and presentation, we will describe the forthcoming BigDAWG release architecture along with a demonstration of polystore applications developed to aid ocean metagenomic researchers explore and analyze their complex datasets.

## II. BigDAWG Release

The initial release of the BigDAWG system will support a number of database engines and islands.

### A. BigDAWG release components

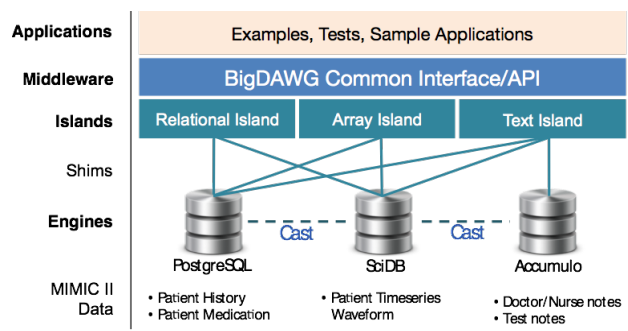

**Figure 1: Notional Architecture of BigDAWG**

The initial release of the BigDAWG system will support three database engines: PostgreSQL (SQL), Apache Accumulo (NoSQL), and SciDB (NewSQL) along with support for relational, array and text islands. A notional architecture of the release is given in Fig 1.

To demonstrate a use case of BigDAWG on real data, the initial release will include scripts to download parts of the publicly-available MIMIC II medical dataset [9] and load it into suitable engines. Patient history data is inserted into PostgreSQL, physiologic waveform data is inserted to SciDB, and free-form text data is inserted into Accumulo. Users will be able to launch the middleware and database engines and issue cross engine queries. We will include a number of example queries and an administrative interface to start, stop and view the status of a BigDAWG cluster.

From a user's perspective, the primary components of the initial system are shown in Fig. 2. A user primarily interacts with the Query Endpoint, which accepts queries, routes them to the Middleware, and responds with results. A detailed description of the Middleware subcomponents are explained in [3]. The Catalog is a PostgreSQL engine containing metadata about the other engines managed by the Middleware. All database engines are containerized and run by Docker, which is in turn managed by an administrative interface.

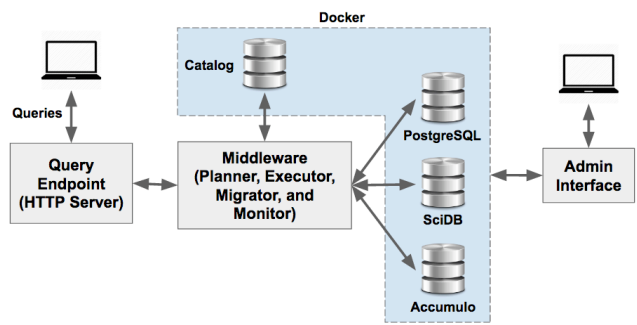

**Figure 2: BigDAWG System Overview**

## B. BigDAWG Containerization

Managing and deploying multiple different databases can be challenging due to platform-specific installation, configuration, startup, and shutdown steps. We are using Docker, which is open-source virtualization software, to simplify and automate the deployment of our database engines and the BigDAWG middleware. This provides a common environment for feature development and allows us to demonstrate the middleware in a way that is reproducible, easily shared, and easily adaptable to new datasets. We selected Docker because it is relatively light to distribute and has shown minimal overhead when compared to other virtualization technologies [10].

For the initial release, a user will download various Docker images from our Docker Hub repository using the `docker pull` command. Then the user can run each image as instantiated containers using the `docker run` command. We have configured each of these containers to connect to a Docker network which allows the containers to communicate with each other and with the host machine. Fig. 3 illustrates the network details.

In order to facilitate future development, we have also built a continuous integration and testing suite that allows errors to be caught as early as possible. We currently use Bitbucket as the code repository and Jenkins as the continuous integration tool. New pushes to the repository and nightly tests ensure that any source modification run as expected. The ability to launch containerized engines during this process greatly simplifies the integration tests. If all integration tests pass, then new Docker images are automatically pushed to our Docker Hub repository so that end-users can pull the latest.

## III. Ocean Metagenomics and BigDAWG

In addition to developing the BigDAWG polystore system, we have continued to find new problems amenable to polystore solutions. The Chisholm Lab at MIT specializes in microbial oceanography and systems biology. The Chisholm Lab works with the GEOTRACES consortium to collect water samples across the globe. These water samples are later analyzed for chemical and hydrographic data and sequenced to determine the relationship between the diversity of cyanobacteria and environmental variables. The Chisholm Lab routinely collects multi-TB of diverse data that consisting of genomic sequences, sensor metadata, hydrographic and chemical data, cruise reports, and streaming data. Chisholm Lab researchers are interested in relationships between communities of cyanobacteria and environmental parameters such as light, temperature and the chemical composition of the seawater. Researchers often struggle with the management of such complex scientific datasets as the variety and volume of data does not easily fit into one single database engine or allow efficient access to all parts of the data.

The BigDAWG team worked with Chisholm Lab researchers to develop a number of applications to help with analysis and exploration of their complex data.

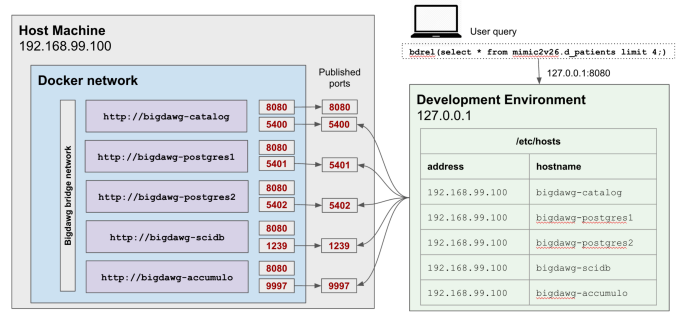

**Figure 3: Container networking and query routing**

## IV. Conclusions and Future Work

The BigDAWG team continues to develop our Polystore solution. Recent developments include application to an ocean metagenomic analysis problem and preparations for an open source release. In this paper, we briefly describe these developments. During the presentation, we intend to give the audience a demonstration of the applications developed for the Chisholm Lab and solicit feedback for our open source release.

## Acknowledgements

This work was supported in part by the Intel Science and Technology Center (ISTC) for Big Data. The authors wish to thank our ISTC Collaborators and members of the Chisholm Lab for their scientific advice.

## References

[1] "Methods to Manage Heterogenous Big Data and Polystore Databases", Workshop at IEEE Big Data 2016 (https://sites.google.com/site/polystoreworkshop/home/program)
[2] Stonebraker, M., & Cetintemel, U. (2005, April). " One size fits all": an idea whose time has come and gone. In *21st International Conference on Data Engineering (ICDE'05)* (pp. 2-11). IEEE.
[3] Gadepally, V., Chen, P., Duggan, J., Elmore, A., Haynes, B., Kepner, J., ... & Stonebraker, M. (2016, December). The BigDAWG polystore system and architecture. In *High Performance Extreme Computing Conference (HPEC), 2016 IEEE* (pp. 1-6). IEEE.
[4] Chen, P., Gadepally, V., & Stonebraker, M. (2016, December). The bigdawg monitoring framework. In *High Performance Extreme Computing Conference (HPEC), 2016 IEEE* (pp. 1-6). IEEE.
[5] Gupta, A. M., Gadepally, V., & Stonebraker, M. (2016, December). Cross-engine query execution in federated database systems. In *High Performance Extreme Computing Conference (HPEC), 2016 IEEE* (pp. 1-6). IEEE.
[6] She, Z., Ravishankar, S., & Duggan, J. (2016, December). BigDAWG polystore query optimization through semantic equivalences. In *High Performance Extreme Computing Conference (HPEC), 2016 IEEE* (pp. 1-6). IEEE.
[7] Dziedzic, A., Elmore, A. J., & Stonebraker, M. (2016, December). Data transformation and migration in polystores. In *High Performance Extreme Computing Conference (HPEC), 2016 IEEE* (pp. 1-6). IEEE.
[8] Elmore, A., Duggan, J., Stonebraker, M., Balazinska, M., Cetintemel, U., Gadepally, V., ... & Madden, S. (2015). A demonstration of the BigDAWG polystore system. *Proceedings of the VLDB Endowment*, 8(12), 1908-1911.
[9] Saeed, M., Villarroel, M., Reisner, A. T., Clifford, G., Lehman, L. W., Moody, G., ... & Mark, R. G. (2011). Multiparameter Intelligent Monitoring in Intensive Care II (MIMIC-II): a public-access intensive care unit database. *Critical care medicine*, 39(5), 952.
[10] Felter, W., Ferreira, A., Rajamony, R., & Rubio, J. (2015, March). An updated performance comparison of virtual machines and linux containers. In *Performance Analysis of Systems and Software (ISPASS), 2015 IEEE International Symposium On* (pp. 171-172). IEEE